\documentclass[paper,notoc]{JHEP3}

\usepackage{epsfig,cite}
\usepackage{amsbsy}
\usepackage{amsmath}
\usepackage{epsfig}
\usepackage{graphicx}

\newcommand{\ep}{\epsilon}
\newcommand{\nn}{\nonumber \\}
\newcommand{\bea}{\begin{eqnarray}}
\newcommand{\eea}{\end{eqnarray}}
\newcommand{\lag}{\ensuremath{{\cal L}}}
\def\beq{\begin{equation}}
\def\eeq{\end{equation}}

\title{\boldmath
Gluon-fusion Higgs production at NNLO for a non-standard Higgs sector 
}

\author{Elisabetta Furlan\\
  Department of Physics, Brookhaven National Laboratory, Upton, NY 11973, USA
  E-mail: \email{efurlan@bnl.gov}}

\abstract{
We  consider an extension of the Standard Model with an arbitrary number of 
heavy quarks having general couplings to the Higgs boson. We construct an effective 
Lagrangian integrating out 
quarks that are heavier than half the mass of the Higgs boson and compute the  
Wilson coefficient for the effective gluon-Higgs vertex through NNLO. 
We apply our result to a composite Higgs model with 
vector-like quarks coupling to the third generation quarks. 
In the heavy quark-mass approximation, we show that the suppression of the leading-order 
cross section with respect to the Standard Model does not depend on the number 
of vector-like multiplets introduced. 
We analyse the effects of QCD and electroweak corrections through three loops, as well
as bottom-quark contributions through two loops.}
\preprint{{ }}

\begin{document}


\section{Introduction}
\label{sec:introduction}

The discovery of the sector responsible for electroweak symmetry 
breaking (EWSB) is the main goal of the LHC. In the minimal description
provided by the Standard Model (SM), the electroweak symmetry is 
broken by an $SU(2)_L$ 
doublet that acquires a vacuum expectation 
value. After EWSB, three of the four degrees of 
freedom associtated to the Higgs doublet provide the longitudinal 
modes of the electroweak gauge bosons, and only one real degree 
of freedom survives: the Higgs boson. In this minimal description, 
the couplings of the Higgs boson to the SM 
particles are fixed by the mass of the particles themselves. 
Indirect experimental bounds point towards a 
relatively light Higgs, with a mass of a few hundred GeV. 
On the other hand, a fundamental scalar receives quadratically 
divergent contributions from radiative corrections.  
A light, fundamental Higgs boson then requires very fine-tuned cancellations between 
the tree-level mass and the higher-order corrections to it. 

New physics scenarios try to address this problem introducing some  
mechanism to protect the mass of the Higgs boson. One of the 
most studied examples is supersymmetry. In supersymmetry new particles are introduced, 
in such a way that their contribution to the self-energy of the Higgs boson counterbalances 
the contribution from the SM particles. 
Another possibility is that the Higgs boson is not a fundamental scalar, but 
a composite state of some new, strongly interacting sector~\cite{Kaplan:1983fs, Kaplan:1983sm}. 
Compositeness would explain the insensitivity of the Higgs boson mass on ultraviolet 
physics, as loop contributions are cut-off at the compositeness scale. 
Such a scale 
is expected to be of the order of a few TeV. Higher values would reintroduce fine-tuning 
problems, while a too low scale would be hard to reconcile with electroweak and flavour physics 
constraints. In this framework, 
the mass of the Higgs boson can be kept naturally light by identifying the Higgs with the pseudo-Goldstone 
boson of some spontaneously broken global symmetry of the new sector, in analogy to 
what happens for the pions in QCD. 
Quarks masses can be generated by mixing the massless fundamental quarks of the SM with 
heavy composite
states of the new sector. This mixing induces a coupling of the SM fermions 
to the Higgs boson. 
As in the SM, only heavy quarks couple significantly to the Higgs 
boson, as they are mainly composite. However, the strength of this coupling 
is not simply proportional 
to the mass, but it depends on the details of the model. 

The introduction of new particles and the modification of the Higgs couplings can
change in a significant way the Higgs-boson phenomenology. At hadron 
colliders, the main mechanism for the production of the Higgs boson is 
gluon fusion. This process is mediated by heavy-particle loops and therefore 
affected both by the introduction of the new quarks, and by the modification 
of the Higgs couplings. 
Within the SM, the gluon-fusion channel has been studied thoroughly. 
The inclusion of next-to-leading order (NLO) ~\cite{Graudenz:1992pv,Spira:1995rr} 
and next-to-next-to-leading order (NNLO)~\cite{Harlander:2002wh, Anastasiou:2002yz, Ravindran:2003um} 
corrections was necessary in order to match the accuracy expected 
by the experiments and to achieve a good converge in the perturbative expansion of 
the cross section in the strong coupling constant $\alpha_s$.
At the LHC, these contributions 
yield an increase of a factor of 2 in the Higgs boson production cross section. 
In composite Higgs models, the LO production cross section 
is expected to be fairly suppressed with respect to the SM 
value~\cite{Falkowski:2007hz, Low:2010mr}. The actual suppression factor depends on 
the details of the model. For an $SO(5) \to SO(4)$ symmetry breaking 
pattern in the strong sector, with composite fermions embedded 
in the fundamental representation of $SO(5)$ and a global symmetry 
breaking scale $f=500 $ GeV, the Higgs boson production 
cross section is expected to be 35\% of the SM value. 
A more detailed study of the Higgs production cross section and 
branching ratios in the different channels was carried out 
in~\cite{Espinosa:2010vn}.
It is interesting to analyse the effect of higher order corrections in 
this class of models, both to check if their effect 
is the same as in the SM and to reach the same accuracy as the SM predictions.

In this paper, we construct an effective Lagrangian integrating out 
the heavy quarks, for which we assume a generic coupling to the 
Higgs boson. We compute the Wilson coefficient for the gluon-Higgs 
effective interaction through NNLO. We apply our result to the 
composite  Higgs model described above, where we introduce  
one or two multiplets of heavy top-partners. 
We include in our study the full bottom-mass dependence 
through NLO~\cite{Spira:1995rr, Anastasiou:2009kn}, 
the two-loop electroweak corrections~\cite{Actis:2008ug} and 
the corresponding three-loop mixed QCD and electroweak 
corrections~\cite{Anastasiou:2008tj}.
These effects are implemented in the program 
\texttt{iHixs}~\cite{iHixs11}. We combine them 
with the NNLO Wilson coefficient that we compute in order to 
obtain the Higgs production cross section through NNLO in 
composite Higgs models. 



\section{The effective Lagrangian}
\label{sec:effective_lagrangian}

We extend the SM quark sector through 
new heavy quarks that transform under the fundamental 
representation of the QCD colour group. 
The number of heavy quarks, including 
the top, is $n_h$. The number of light quarks 
is $n_l$. We take the light quarks to be massless and  not to 
couple to the Higgs boson.
We assume instead an arbitrary coupling 
$Y_i$ of the heavy quarks to the Higgs boson. In 
the SM, $Y_i = m_i/v$, where $v \approx $ 246 GeV is the vacuum expectation value of 
the Higgs boson and $m_i$ is the mass of the quark. 
The Lagrangian describing this model is
\begin{equation}
\label{eq:lag_full}
  {\cal L}
  =
      {\cal L}^{n_l}_{QCD}
  + \sum_{i=1}^{n_h}  \bar{\psi}_i 
  	  \left(  
  			i D \!\!\!\!\slash - m_i 
  	   \right) 
       \psi_i
    + {\cal L}_Y
    \quad {\rm ,\;with} \quad
    {\cal L}_Y
    =
    - H
    	\sum_{i=1}^{n_h}  
    				Y_i  \bar{\psi}_i  \psi_i
       \; .
\end{equation}
Here $D_{\mu}$ is the covariant derivative in the fundamental representation 
of the colour group
and $ {\cal L}^{n_l}_{QCD}$ is the QCD Lagrangian with only the 
$n_l$ flavours of light quarks. 

We focus on the Higgs production from gluon fusion mediated by loops of 
heavy quarks.
When the particles that couple to the Higgs boson are heavier than half the 
Higgs boson mass, we can integrate them out.
In this limit, we can replace the original Lagrangian (\ref{eq:lag_full}) 
with an effective Lagrangian
\begin{equation}
	{\cal L}^{eff} 
	=
	   {\cal L}_{QCD}^{eff,n_l}
     - C_1 \,H \,{\cal O}_1
     \;.
\label{eq:Leff}
\end{equation}
The dimension-four local operator 
${\cal O}_1$ reads~\cite{Spiridonov:1984br} 
\begin{equation}
	{\cal O}_1 
	= 
	\frac{1}{4} \,{G'}^a_{\mu\nu} {G'}^{a\mu\nu} 
	\; ,
\end{equation}
where ${G'}^a_{\mu\nu}$ is the field strength tensor 
in the effective theory, and $C_1$ is the corresponding 
Wilson coefficient~\cite{Wilson:1969zs}. 
The effective Lagrangian 
$  {\cal L}_{QCD}^{eff,n_l}$ appearing in Eq.~(\ref{eq:Leff}) 
describes the interactions among the light  
degrees of freedom. It has the same form as
$  {\cal L}_{QCD}^{n_l}$, but with different coupling and field normalizations 
to account for the missing contributions from heavy quarks loops.  
The parameters in the effective theory are related to the parameters in 
the full theory through multiplicative decoupling constants $\zeta_i$. 
The derivation of the decoupling constants is reviewed in~\cite{Steinhauser:2002rq,Anastasiou:2010bt}.



\section{Details of the calculation}
\label{sec:details_of_calculation}


We compute the Wilson coefficient $C_1$ up to three
loops. 
We start from the bare amplitude
${\cal M }^0_{gg \to H}$ 
for the process $g g \to H$ in the full theory, 
\begin{equation}
\label{eq:M0ggh}
	{\cal M }^0_{gg \to H} 
	\equiv 
	{\cal M}^{0,a_1 a_2}_{\mu_1\mu_2}(p_1, p_2) 
		\epsilon^{\mu_1}_{a_1} \epsilon^{\mu_2}_{a_2}
	\;.
\end{equation}
Here $p_1$ and $p_2$ are the momenta of the two gluons 
and $\epsilon^{\mu_1}_{a_1}$, $\epsilon^{\mu_2}_{a_2}$
are their polarizations. 
This amplitude is related to the bare Wilson coefficient $C^0_1$ by~\cite{Steinhauser:2002rq}
\begin{equation}
	\zeta^0_3 C^0_1
 	=
	\frac{\delta^{a_1 a_2}
			\left( g^{\mu_1\mu_2} (p_1\cdot p_2) - p_1^{\mu_2} p_2^{\mu_1} \right)
			}
			{ (N^2-1) (d-2) {(p_1\cdot p_2)^2} } 
	{\cal M}^{0,a_1 a_2}_{\mu_1\mu_2}(p_1, p_2)
   \bigr\rvert_{p_1 = p_2 = 0}
   \;,
\label{eq:C01}
\end{equation}
with $N$ the number of colours and $d=4-2\epsilon$ the dimension of
space-time. The superscript $``0"$ denotes bare quantities. 
The factor $\zeta^0_3$ is the bare decoupling
coefficient 
that relates the bare gluon field 
${G'}^{0,a}_{\mu} $ 
in the effective theory to the bare gluon field 
${G}^{0,a}_{\mu} $ 
in the full theory, 
\begin{equation}
\label{eq:def_zeta03}
	{G'}^{0,a}_{\mu} 
	= 
	\sqrt{\zeta^0_3} \,{G}^{0,a}_{\mu} 
	\; .
\end{equation}

We  generate  the Feynman diagrams  ${\cal F}$ 
for the amplitude through three loops using QGRAF~\cite{Nogueira:1991ex}. 
Diagrams containing two different heavy mass
scales appear for the first time at the three-loop order. 
We then expand all diagrams in the external momenta $p_1,p_2$ 
by applying the following  differential operator~\cite{Fleischer:1994ef} 
to their integrand: 
\begin{equation}
	{\cal D} {\cal F} 
	= 
	\sum_{n=0}^\infty (p_1 \cdot p_2)^n \left[ {\cal D}_n {\cal F} \right]_{p_1 = p_2 =0}, 
\end{equation}
with 
\begin{equation}
{\cal D}_0 =1, 
\quad 
{\cal D}_1 =\frac{1}{d} \Box_{12},
\qquad
 {\cal D}_2 =-\frac{1}{2(d-1)d(d+2)} \left\{ 
\Box_{11} \Box_{22} - d\; \Box_{12}^2
\right\}, 
\end{equation}
and $ \Box_{ij} \equiv g^{\mu \nu}\frac{\partial^2}{\partial p_i^\mu \partial p_j^\nu}$. 
Differential operators of higher orders are not needed for the expansion 
in the external momenta at leading order.

After Taylor expansion, all the Feynman diagrams are expressed in terms 
of one-, two- and three-loop vacuum bubbles. 
We  reduce these integrals to a set of five master integrals 
using the algorithm of Laporta~\cite{Laporta:2001dd} and the
program AIR~\cite{Anastasiou:2004vj}. 
The topologies and the master integrals that we find are the same 
as in the calculation of the Wilson coefficient for an arbitrary 
number of heavy quarks with top-like Yukawa interactions of Ref.~\cite{Anastasiou:2010bt}, 
\begin{eqnarray}
	I_{1}
	& = & 
	\int \!\! \frac{d^dk_1}{i \pi^{d/2}}
				 \frac{1}{{\cal P}_1} 
	= - \left( m_i^2\right)^{1-\epsilon} \Gamma(-1+\epsilon) \; , \\
	I_{2}
	& = & 
	\int \!\! \frac{d^d k_1 d^d k_2 d^dk_3}{(i \pi^{d/2})^3}
		\frac{  1  }{  {\cal P}_1 {\cal P}_3 {\cal P}_5  {\cal P}_6}
	\nonumber \\
	&=& \left( m_i^2 \right)^{2-3\epsilon} 
 						\frac{ \Gamma^2(1-\epsilon) \Gamma(\epsilon) \Gamma^2(-1+2\epsilon) \Gamma(-2+3\epsilon)}
 								{\Gamma(2-\epsilon)\Gamma(-2+4\epsilon)}
		\;, \\ 
	I_{3}
	& = &
	\int \!\! \frac{d^d k_1 d^d k_2 d^dk_3}{(i \pi^{d/2})^3}
	\frac{  1  }{  {\cal P}_1  {\cal P}_2  {\cal P}_3 {\cal P}_4 }
	\;, \\ 
	I_{4}
	& = &
	\int \!\! \frac{d^d k_1 d^d k_2 d^dk_3}{(i \pi^{d/2})^3}
	\frac{  1  }{  {\cal P}_1 \tilde{{\cal P}}_2  \tilde{{\cal P}}_3 {\cal P}_4 }
	\;, \\ 
	I_{5}
	& = &
	\int \!\! \frac{d^d k_1 d^d k_2 d^dk_3}{(i \pi^{d/2})^3}
		\frac{  1  }{  {\cal P}_1^2  \tilde{{\cal P}}_2 \tilde{{\cal P}}_3 {\cal P}_4 }
	\; ,
\end{eqnarray}
with  
\begin{equation}
\begin{array}{ccll}
	{\cal P}_{1} &=& k_1^2- m_i^2 \;,  & \\
	{\cal P}_{2} &=& k_2^2- m_i^2 \;,  & \tilde{{\cal P}}_{2} = k_2^2- m_{j}^2 \;, \\
	{\cal P}_{3} &=& k_3^2- m_i^2 \;,  & \tilde{{\cal P}}_{3} = k_3^2- m_{j}^2 \;, \\
	{\cal P}_{4} &=& (k_1 -k_2 + k_3)^2 - m_i^2 \;, \quad & \\
	{\cal P}_{5} &=& (k_1 -k_2)^2 \;, & \\ 
	{\cal P}_{6} &=& (k_2 - k_3)^2 \;,  & 
\end{array}
\end{equation}
The last two master integrals contain two heavy quarks of different 
mass, $m_i$ and $m_{j}$. They have been computed in~\cite{Bekavac:2009gz}. 
The one-scale master integrals can be found for example in~\cite{Steinhauser:2000ry}. 

For completeness, we report the Wilson coefficient in terms of the 
bare parameters in the full theory, 
\bea
\label{eq:bare_WC_genyuk}
	\zeta^0_3 C^0_1
 	&=&
 	\frac{1}{3}
 	\left( \frac{\alpha_s^0 S_{\epsilon}}{\pi} \right) 
 	\left[
 			-\Upsilon_0^0	+\epsilon \left[	\Upsilon_0^0	 + 2 \Upsilon_1^0	 	\right]	
 			- 2 \epsilon^2 \left(	\Upsilon_2^0	+ \Upsilon_1^0	+	\Upsilon_0^0\frac{\pi^2}{24}	\right)	+ {\cal O}(\epsilon^3)
 	\right]\nn
 	&&
 	+
 	\left( 	\frac{  \alpha_s^0 S_{\epsilon}  }{  \pi  } 	\right)^2
	\left[	
		-\frac{\Upsilon_0^0 }{ 4 }	+ \epsilon \left(	\Upsilon_1^0 + \frac{31}{36} \Upsilon_0^0	\right)
		+ {\cal O}(\epsilon^2)
	\right]	\nn
	&&
	+
	\left( 	\frac{\alpha_s^0 S_{\epsilon}}{\pi} 		\right)^3 
	\left\{
			-\frac{ n_h^2 }{32 \epsilon^2} 	\Upsilon_0^0	
			+\frac{1}{\epsilon} 
			   	\left[
			   			\Upsilon_0^0	\left(\frac{89}{576} n_h -\frac{13}{24} - \frac{5}{144} n_l + \frac{1}{8} L_1^0		\right)
			   			+\frac{n_h}{16}	\Upsilon_1^0	
			 	 \right]
			 \right.	 \nn
	&&
	\phantom{-}
	 +\Upsilon_0^0	\left[	\frac{1171}{576} +\frac{103}{864} n_l - n_h \left( \frac{ 1051 }{ 3456 } +\frac{\pi^2}{128}\right)
			 - \frac{89}{96} L_1^0 -\frac{3}{8} L_2^0 \right] 
	\nn	&&  
	\phantom{-}
			+\left(\frac{13}{4}+\frac{5}{24} n_l\right) \Upsilon_1^0
			-	\frac{3}{16} n_h \Upsilon_2^0
	\nn
%
	& & 
	\phantom{-}
	- \sum_{\substack{1 \leq i < n_h \\ i<j \leq n_h}} 
		\left[	(y^0_i-y^0_j)
			\left(
				\frac{19}{128} 
				\bigg(\frac{(m_i^0)^2}{(m_j^0)^2}-\frac{(m_j^0)^2}{(m_i^0)^2}\bigg)
			 \right. \right. \nn
	& & 	\phantom{---}
				+ \bigg(
						\frac{19}{128} \frac{(m_i^0)^2}{(m_j^0)^2}+\frac{19}{128} \frac{(m_j^0)^2}{(m_i^0)^2}+\frac{43}{96}
					\bigg)	
				\log \bigg(\frac{m_i^0}{m_j^0}\bigg)
\nonumber 
\eea 
\bea
	& & 	\phantom{-------}
	\left.
			+\frac{19}{256} \frac{(m_i^0)^6+(m_j^0)^6}{ (m_i^0)^2 (m_j^0)^2 \left( (m_i^0)^2-(m_j^0)^2\right)} \log ^2 \bigg(\frac{m_i^0}{m_j^0}\bigg)
		\right)	\nn
	& & 
	\phantom{-----}
		-\log ^2 \bigg(\frac{m^0_i}{m^0_j}\bigg)
		\left(
			\frac{73}{768} \left(y^0_i + y^0_j \right)
			+ \frac{23}{384}\frac{y^0_i (m_i^0)^2 - y^0_j (m_j^0)^2}{(m_i^0)^2-(m_j^0)^2}
			\right.\nn
	& & \left. 
	\phantom{-------}
			+\left( (m_i^0)^2-(m_j^0)^2\right) \frac{19 (m_i^0)^4+24 (m_i^0)^2 (m_j^0)^2+19 (m_j^0)^4}{512 (m_i^0)^3 (m_j^0)^3} 
			\right.\nn
	& & \left. 
	\phantom{---------}
			\cdot \bigg(
				y^0_j \log \Big(\frac{m^0_j - m^0_i}{m^0_j+m^0_i} \Big)
				-y^0_i \log \Big(\frac{m^0_i - m^0_j}{m^0_i+m^0_j} \Big)
			\bigg)
		\right) \nn
& & \left. \left.
			\phantom{-----}
		-\frac{19 (m_i^0)^6+5 (m_i^0)^4 (m_j^0)^2-5 (m_i^0)^2 (m_j^0)^4-19 (m_j^0)^6}{1024 (m_i^0)^3 (m_j^0)^3}
		\right. \right. \nn
& & \left. \left.
		\phantom{------} \cdot
		\left(8 y^0_i \text{Li}_3 \bigg(\frac{m^0_j}{m^0_i} \bigg) -8 y^0_j \text{Li}_3 \bigg(\frac{m^0_i}{m^0_j} \bigg)
			-y^0_i \text{Li}_3 \bigg(\frac{(m_j^0)^2}{(m_i^0)^2} \bigg)+y^0_j \text{Li}_3 \bigg(\frac{(m_i^0)^2}{(m_j^0)^2} \bigg)
		\right. \right. \right. \nn
& & \left. \left. \left. 
		\phantom{---}
		-2 \log\! \bigg(\frac{m^0_i}{m^0_j} \!\bigg) \bigg( y^0_i \text{Li}_2\! \Big(\frac{(m_j^0)^2}{(m_i^0)^2} \!\Big)
			+y^0_j \text{Li}_2\! \Big(\frac{(m_i^0)^2}{(m_j^0)^2} \!\Big)-4 y^0_i \text{Li}_2\! \Big(\frac{m^0_j}{m^0_i} \Big)
			-4 y^0_j \text{Li}_2\! \Big(\frac{m^0_i}{m^0_j} \Big)\!\bigg)
			\right)
	\right]
	\right\}.\nn
\eea
Here we introduced the notation
\bea
\label{eq:def_Ypsilon0}
	S_{\epsilon} & = &	 e^{-\epsilon \gamma_E}  \left(  4 \pi  \right)^\epsilon  \quad , \nn
	L_1^0 &=& \sum_{i=1}^{n_h} \log(m_i^0)
	\quad , \quad
	L_{2}^0 = \sum_{i=1}^{n_h} \log^2(m_i^0)
	\quad , \nn
\Upsilon_0^0  &=&  \sum_{i=1}^{n_h}  y_i^0  \;,\quad 
\Upsilon_1^0  = 	\sum_{i=1}^{n_h}   y_i^0   \log(m_i^0)  \;,\quad 
\Upsilon_2^0  =  \sum_{i=1}^{n_h}   y_i^0  \log^2(m_i^0)	\;,
\eea
with 
\beq
	    y_i^0 = \frac{ Y_i^0 }{ m_i^0 }	 \; .
\eeq	    
This expression reproduces the result of Ref.~\cite{Anastasiou:2010bt} 
when $Y_i^0 = \frac{m_i^0}{v}$. 
%


 
\subsection{Decoupling and renormalization}

The RHS of Eq.~(\ref{eq:C01}) is expressed in terms 
of the bare coupling constant $\alpha_s^0$ in the full theory and of 
the bare masses and Yukawa couplings of the heavy quarks, $m_i^0$ 
and $Y_i^0$; $C_1^0 = C_1^0( \alpha_s^0, m_i^0, Y_i^0) $.
%
%
The bare strong coupling in the full theory is related to the 
bare strong coupling in the effective theory $\alpha_s^{'0}$
by the decoupling constant $\zeta_g^0$
~\cite{Steinhauser:2002rq, Chetyrkin:1997un}, 
\begin{equation}
	\label{eq:decouple_bare}
	\alpha_s^{'0}   =   (\zeta_g^0)^2 \alpha_s^{0} \quad \;.
\end{equation}
The decoupling constants of the strong coupling and of the gluon field
(Eq.~(\ref{eq:def_zeta03})) in the presence of an arbitrary number of 
heavy quarks have been derived in~\cite{Anastasiou:2010bt}, 

\begin{eqnarray}
	\zeta_g^0
	&=&
	   1
	+ \left( \frac{\alpha_s^0 S_{\epsilon}}{\pi} \right) 
		\left[
			-\frac{n_h}{12 \epsilon}
			+\frac{1}{6} L_1^0
			- \epsilon \left(  n_h \frac{ \pi ^2 }{144}  
							+ \frac{1}{6}  L_2^0
							\right)
		\right]
	\nonumber \\
	& & \phantom{1}
	+ \left( \frac{\alpha_s^0 S_{\epsilon}}{\pi} \right)^2 
		\left\{
			  \frac{n_h^2}{96 \epsilon^2}
			- \frac{n_h}{ 24 \epsilon}
				\left[ \frac{3}{4}  +  L_1^0 \right]
			+\frac{1}{8}  L_1^0
			+ \frac{1}{24} \left(  L_1^0\right)^2 
			+\frac{n_h}{24}
				\left[  \frac{  11  }{  6  }  + L_2^0 \right] 
			+ n_h^2 \frac{\pi^2 }{576}
		\right\} \;, \nn
 \zeta^0_3
 &=&
 1
 +
 \left\{
	\left( \frac{\alpha_s^0 S_{\ep}}{\pi} \right) 
	\left[
			   \frac{n_h}{6 \ep}
			- \frac{1}{3} L_1^0
			+	\ep \left(\frac{\pi^2}{72} n_h + \frac{1}{3 } L_2^0 \right)
	\right]
	 \right.\nn
	&&
	\left. \phantom{-}
	+ \left(  \frac{  \alpha_s^0 S_{\ep}  }{  \pi  } \right)^2 
	\left[
			   \frac{3}{32 \ep^2} n_h
			 -\frac{n_h+24 L_1^0}{64 \ep}
			+\frac{3}{4} L_2^0
			+\frac{1}{16} L_1^0
			+ n_h \left(\frac{91}{1152}
			+\frac{\pi^2}{64} \right)
	\right]	
\right\} \;.
\end{eqnarray}


After decoupling, the bare Wilson coefficient is a 
function of the bare parameters in the effective theory and of the bare masses 
and Yukawa couplings of the 
heavy quarks in the full theory, 
$C_1^0 =C_1^0( \alpha_s^{'0}, m_i^{0}, Y_i^{0}) $. 
The bare parameters are related to the renormalized ones through multiplicative 
renormalization constants $Z_i$ as
\begin{eqnarray}
	\label{eq:renormalization_effective}
	\alpha_s^{'0}   &=&  \mu^{2 \epsilon} Z_{\alpha}' \alpha_s'(\mu) \quad , \quad
	\phantom{m_l^{'0}   =   Z'_{m_l} m'_l(\mu) \;,} \\
	\label{eq:renormalization_full}
	m_i^{0}   &=&   Z_{m_i} m_i(\mu)
	\quad\;\,, \quad 
	 Y_i^{0}   =   Z_{Y_i} Y_i(\mu)\;.
\end{eqnarray}
The primed parameters in Eq.~({\ref{eq:renormalization_effective})
are in the effective theory 
and the unprimed parameters in 
Eqs.~({\ref{eq:renormalization_full}) are in the full theory. 
In the $\overline{MS}$ scheme the strong coupling renormalization constant 
is 
\begin{equation}
\label{eq:Zalpha}
	Z'_{\alpha}
	= 
	   1
	- \frac{  \alpha'_s ( \mu )  }{  \pi  }
		\frac{  \beta'_0  }{  \epsilon  }
	+ \left(
			\frac{  \alpha'_s ( \mu )  }{  \pi  }
		\right)^2
		\left(
			\frac{  \beta_0^{'2}  }{  \epsilon^2  }
			-
			\frac{  \beta'_1  }{  2 \epsilon  }
		\right)
	\;,
\end{equation}
where $\beta'_0$, $\beta'_1$ denote the first two coefficients of the $\beta$ function in the
light-flavour theory,
\begin{equation}
	\beta'_0 
	=  
	\frac{ 1 }{ 4 }
	\left(  11  -  \frac{ 2 }{ 3 } n_l  \right)
	\quad , \quad 
	\beta'_1 
	= 
	\frac{ 1 }{ 16 }
	\left(  102  - \frac{ 38 }{ 3 } n_l  \right)
	\;.
\end{equation}
The mass renormalization constant reads~\cite{Tarrach:1980up}
\begin{equation}
\label{eq:Zmq}
	Z_{m_i}
	=
	   1
	- \frac{ \alpha_s(\mu) }{ \pi  }
		\frac{ 1 }{ \epsilon }
	+\left(      \frac{ \alpha_s(\mu) }{ \pi }     \right)^2
		\left[
			\frac{ 1 }{ \epsilon^2 }
			\left(
					\frac{ 45 - 2 n_f }{ 24 } 
			\right)
			+
			\frac{ 1 }{ \epsilon }
			\left(
					- \frac{ 101 }{ 48 }	+	\frac{ 5 }{ 72 } n_f
			\right)
	\right]
 \;,
\end{equation}
where $n_f$ is the number of active flavours. In the 
effective theory $n_f = n_l$ and in the full theory $n_f = n_l + n_h$. 
%
%
Finally, the Yukawa coupling $Y_{i}^0$ 
renormalizes as the mass, 
\begin{equation}
 	Z_{Y_i} = Z_{m_i} \;.
\end{equation}
We review here the proof of this relation, following 
Ref.~\cite{Chetyrkin:1994js}. 

Let us consider the bare scalar current 
\beq
	j^0(x) = \overline{\psi}^{\, 0}_i(x) \psi^0_i(x) \;.
\eeq
We relate it to the renormalized scalar current $j(x)$ through 
a multiplicative renormalization constant $Z_J$, 
\beq
	j^0(x) = Z_J j(x)\;.
\eeq

Let us recall that the bare and renormalized two-point Green functions for the fermions 
with an insertion of the scalar current are defined 
respectively as 

\bea
	G^{0,(2)}(p_1, p_2) &=& 
	i^2 \int \!\! dx_1 dx_2 \, e^{i (p_1 x_1-p_2 x_2)} 
	\langle T \psi_i^0 (x_1) j^0(0) \overline{\psi}_i^{\, 0}(x_2) \rangle
\nn &\equiv &
	i^2 \int \!\! dx_1 dx_2 \, e^{i (p_1 x_1-p_2 x_2)} 
	\left\{
		\int [ {\cal D} \Phi^0] e^{i \! \int \!\! dx \, {\lag}^0} j^0(0) \psi_i^0 (x_1) \overline{\psi}_i^{\, 0}(x_2)
	\right\}
	 \;,\nn
	\label{eq:g02j}
\\
	G^{(2)}(p_1, p_2) &=& 
	i^2 \int \!\! dx_1 dx_2 \, e^{i (p_1 x_1-p_2 x_2)} 
	\langle T \psi_i (x_1) j(0) \overline{\psi}_i(x_2) \rangle
\nn &\equiv &
	i^2 \int\!\! dx_1 dx_2 \, e^{i (p_1 x_1-p_2 x_2)} 
	\left\{
		\int [ {\cal D} \Phi] e^{i \! \int \!\! dx \, {\lag}\,} j(0) \psi_i (x_1) \overline{\psi}_i(x_2)
	\right\}
	 \;.
	 \label{eq:g2j}
\eea 
Here $\lag^0, \lag\,$ are respectively the bare and renormalized QCD Lagrangian and 
$ \int [ {\cal D} \Phi^0]$, $\int [ {\cal D} \Phi] $ denote the functional integral over the bare and renormalized 
fields appearing in them. 
From the second line of both these equalities we see that $G^{0,(2)}$, $G^{(2)}$ are related 
to the bare and renormalized quark propagators 
\bea
	S^0(p) & \equiv & i \int \! \! d^4 x\,  e^{ipx} \langle  T \psi^0_i (x) \overline{\psi}^{\, 0}_i(0)  \rangle 
		= i \int \!\! dx \, e^{i p x} 
	\left\{
		\int [ {\cal D} \Phi^0] e^{i \! \int \!\! dy \, {\lag}^0}  \psi_i^0 (x) \overline{\psi}_i^{\, 0}(0)
	\right\} \;, \nn
	S(p) &\equiv & i \int \!\! d^4 x\, e^{ipx} \langle  T \psi_i (x) \overline{\psi}_i(0)  \rangle 
	= i  \int\!\! dx \, e^{i p x} 
	\left\{
		\int [ {\cal D} \Phi] e^{i \! \int \!\! dy \, {\lag}\,}  \psi_i (x) \overline{\psi}_i(0)
	\right\}\;,
\eea
as 
\beq
	G^{0,(2)}(p, p) = - \frac{\partial }{\partial m_i^0 } S^0 (p) \;, \quad
	G^{(2)}(p, p) = - \frac{\partial }{\partial m_i } S (p) \;.
\eeq
Therefore 
\beq
	G^{0,(2)} = \frac{Z_2}{Z_{m_i}} G^{(2)}\;,
\eeq
where $Z_2$ is the renormalization constant of the quark wavefunction,
\beq
	\psi_i^0 = \sqrt{ Z_2 } \psi_i \;.
\eeq
On the other hand, comparing Eqs.~(\ref{eq:g02j}) and (\ref{eq:g2j}) we have 
\beq 
	G^{0,(2)} = Z_2 Z_J G^{(2)}\;,
\eeq
and therefore
\beq 
	Z_J  =Z_{m_i}^{-1} \;.
\eeq
The bare Yukawa Lagrangian $\lag_Y^0$ has the same form of 
 $\lag_Y$ (Eq.~(\ref{eq:lag_full})), but it contains bare fields and 
 Yukawa couplings. Treating the Higgs field as an external static
 field, we can use the result for the renormalization of the scalar
 current $j^0(x)$ to relate the bare and renormalized Yukawa 
 Lagrangians. We obtain
\beq 
 	Z_{Y_i} = Z_J^{-1} = Z_{m_i}\;.
\eeq


We finally renormalize the bare Wilson coefficient $C_1^0( \alpha'_s, m_i, Y_i )$ 
using~\cite{Spiridonov:1984br,Spiridonov:1988md,Chetyrkin:1996ke}
\begin{eqnarray}
	C_1
	& = &
	\frac{  1  }{  1 + \alpha'_s(\mu) 
	\frac{  \partial  } {  \partial \alpha'_s(\mu)  }
	\log Z'_{\alpha}}	C^0_1
	=	\left[
		1	+	\frac{  \alpha'_s(\mu)  }{  \pi  }
				\frac{  \beta'_0  }{  \epsilon  }
			+	\left(
					\frac{  \alpha'_s(\mu)  }{  \pi  }
				\right)^2
				\frac{  \beta'_1  }{  \epsilon  }
		\right]		C^0_1	\;.
\end{eqnarray}
%
%
Our final result for the renormalized Wilson coefficient reads
\begin{eqnarray}
\label{eq:WCfinal}
C_1 
&=& 
	-\frac{1}{3} \Upsilon_0
	- \frac{11}{12} \frac{\alpha_s'(\mu)}{\pi} \Upsilon_0
	-\frac{1}{3} \left(\frac{\alpha_s'(\mu)}{\pi} \right)^2 \left\{
		-n_l \left(
		\frac{67}{96} \Upsilon_0 +\frac{2}{3} \Upsilon_1\right)
	\right. \nn
	& & \left.
	+ \Upsilon_0 \left(
		\frac{1877}{192}-\frac{77}{576} n_h 
			+ \frac{113}{96} L_1 + \frac{ 3 }{ 8 } L_2
		\right)
	- \Upsilon_1 \left(
		\frac{19}{8} + \frac{113}{96} n_h 
		+ \frac{3}{4} L_1 \right)
	+\frac{3}{8} n_h \Upsilon_2
	\right. \nonumber\\
	& & \left.
	+ \sum_{\substack{1 \leq i < n_h \\ i<j \leq n_h}} \left[
			(y_i-y_j) \left(
				\frac{57}{128} \bigg(\frac{m_i^2}{m_j^2}-\frac{m_j^2}{m_i^2} \bigg)
			+\bigg(
				\frac{57}{128} \frac{m_i^2}{m_j^2}+\frac{57}{128} \frac{m_j^2}{m_i^2}+\frac{43}{32}
			\bigg) \log \left(\frac{m_i}{m_j}\right)
			\right. \right. \right. \nn
	& & \left. \left. \left.
	\phantom{-}
			+\frac{57}{256} \frac{m_i^6+m_j^6}{ m_i^2 m_j^2 \left(m_i^2-m_j^2\right)} \log ^2\left(\frac{m_i}{m_j}\right)
		\right)
		\right. \right. \nn
& & \left. \left.
	\phantom{-}
		-\log ^2\left(\frac{m_i}{m_j}\right)
		\left(
			\frac{73}{256} \left(y_i + y_j \right)
			+ \frac{23}{128}\frac{y_i m_i^2 - y_j m_j^2}{m_i^2-m_j^2}
			\right. \right. \right. \nn
& & \left. \left. \left.
	\phantom{-}
			+\frac{3}{512}\! \left(m_i^2-m_j^2\right) \frac{19 m_i^4+24 m_i^2 m_j^2+19 m_j^4}{ m_i^3 m_j^3} 
			\bigg(\!
				y_j \log\! \Big(\frac{m_j - m_i}{m_j+m_i} \Big)
				-y_i \log\! \Big(\frac{m_i - m_j}{m_i+m_j} \Big)
			\! \bigg)
		\right)
		\right. \right. \nn
& & \left. \left.
	\phantom{-}
		-\frac{3}{1024}\frac{19 m_i^6+5 m_i^4 m_j^2-5 m_i^2 m_j^4-19 m_j^6}{ m_i^3 m_j^3}
		\right. \right. \nn
& & \left. \left.
\phantom{--}
		\cdot \left(8 y_i \text{Li}_3 \Big(\frac{m_j}{m_i} \Big) -8 y_j \text{Li}_3 \Big(\frac{m_i}{m_j} \Big)
			-y_i \text{Li}_3 \Big(\frac{m_j^2}{m_i^2} \Big)+y_j \text{Li}_3 \Big(\frac{m_i^2}{m_j^2} \Big)
		\right. \right. \right. \nn
		& & \left. \left. \left. 
\phantom{--}
		-2 \log \Big(\frac{m_i}{m_j} \Big) \bigg(y_i \text{Li}_2 \Big( \frac{m_j^2}{m_i^2} \Big)+y_j \text{Li}_2 \Big(\frac{m_i^2}{m_j^2} \Big)
		-4 y_i \text{Li}_2 \Big(\frac{m_j}{m_i} \Big)-4 y_j \text{Li}_2 \Big(\frac{m_i}{m_j} \Big) \bigg)\right)
	\right]
	\right\}\,.\nn
\end{eqnarray}
Similarly to Eq.~(\ref{eq:def_Ypsilon0}), we defined
\bea
\label{eq:def_Ypsilon}
	L_1 &=& \sum_{i=1}^{n_h} \log \left(\frac{m_i}{\mu}\right)
	\quad , \quad
	L_{2} = \sum_{i=1}^{n_h} \log^2\left(\frac{m_i}{\mu}\right)
	\quad , \nn
	\Upsilon_0  &=&  \sum_{i=1}^{n_h}  y_i  \;,\quad 
	\Upsilon_1  = 	\sum_{i=1}^{n_h}   y_i   \log \left(\frac{m_i}{\mu}\right) \;,\quad 
	\Upsilon_2  =  \sum_{i=1}^{n_h}   y_i  \log^2\left(\frac{m_i}{\mu}\right)	\;,
\eea
with 
\beq
	    y_i = \frac{ Y_i }{ m_i }	 \quad .
\eeq	    
For $Y_i = \frac{m_i}{v}$ we recover the Wilson coefficient for an arbitrary number of 
heavy quarks with Standard Model-like 
Yukawa interactions of~\cite{Anastasiou:2010bt}.



\section{Composite Higgs models}
\label{sec:composite_higgs_models}

In this Section we review briefly the composite Higgs model of 
Refs.~\cite{Gillioz:2008hs, Anastasiou:2009rv}.


\subsection{The Higgs sector}
\label{sec:higgs_sector}

At low energy, the strong sector responsible for EWSB in composite Higgs scenarios 
can be described by a non-linear sigma model. Such description allows to decouple 
the light pseudo-Goldstone boson from other heavy composites of the new sector. 
We will consider the global symmetry breaking pattern 
$SO(5) \to SO(4)$, 
as this is the minimal pattern that includes custodial symmetry. 
An additional $U(1)_X$ symmetry is required 
in order to generate the correct Weinberg angle. The SM
electroweak group $SU(2)_L \times U(1)_Y$ is embedded 
into $SO(4)\times U(1)_X \sim SU(2)_L\times SU(2)_R \times U(1)_X$ 
so that the hypercharge is given by $Y = T^{3}_{R} + X$~\cite{Agashe:2004rs, 
Contino:2006qr}.
We denote by $f$ the scale at which the global symmetry is broken. 
This scale is assumed to be larger than the 
EWSB scale $v \approx 246$~GeV. Too large values of $f$ would introduce a 
substantial fine-tuning of the model~\cite{Barbieri:2007bh}; on the other hand, 
if the scale of new physics is too low, large contributions to electroweak parameters 
and flavour physics are expected. We set $f=500$~GeV, which corresponds to a 
~$\sim 10\%$ fine-tuning~\cite{Barbieri:2007bh}.

The effective theory becomes strongly coupled at a scale $\sim 4 \pi f$. 
Above this energy, 
a more fundamental description needs to be introduced. If the coupling 
constant of the strong sector $g_\rho$ is not maximal, i.e. $g_\rho < 4 \pi$, 
hadronic bound states appear below the strong-coupling scale. The typical mass of 
these resonances is $m_\rho \simeq g_\rho f$ and acts as a lower, effective 
cut-off for the effective-theory description. Thorough this work we will assume the presence of 
such weakly-coupled new states 
and take as cut-off $\Lambda = m_\rho \simeq 2 \pi f$.

The $SO(5) \rightarrow SO(4)$ breaking 
is realized through a scalar field $\Sigma(x)$ subject to the constraint 
\beq
	\Sigma^2(x) = 1 \;.
\eeq
In the non-linear representation 
\beq
\label{eq:Simga10000}
		\Sigma(x) = \Sigma_0 e^{\Pi(x)/f} \quad, \quad\quad
		\Pi(x) = -i T^{\hat{a}} h^{\hat{a}}(x) \sqrt{2} \;,
\eeq
where $\Sigma_0 = (0,0,0,0,1)$ 
is the vacuum state that preserves $SO(4)$, 
$T^{\hat{a}}$ are the four broken generators 
and $h^{\hat{a}}$ the corresponding Goldstone bosons. 
Eq.~(\ref{eq:Simga10000}) can be rewritten as 
\beq
		\Sigma 
		= \left(
				\frac{	h^{ \hat{a} }	}{	h	}
				\sin \frac{ h }{ f }
				,
				\cos \frac{ h }{ f }
			\right)
		\equiv
			\left(
				\vec{\Sigma},
				\Sigma_5
			\right)
		\;,
\label{eq:sigma_expanded}
\eeq
where $\vec{h} = (\phi^{0*}, - \phi^{-}, \phi^{+}, \phi^{0}) $
transforms under the fundamental representation of 
$ SO(4)$ and 
$ h 	= 	\sqrt{ \left( h^{ \hat{a} } \right)^2} $. 
The SM Higgs doublets with hypercharge +1/2 and -1/2
are respectively 
$
		\Phi = \left(		\phi^{+} , \phi^0 	\right)^T
$
and 
$
		\tilde{\Phi} = i \sigma^2 \Phi^{*}
$. 
The requirement 
\beq
	m_W^2 = \frac{ g^2 f^2 }{ 4 } \sin^2 \left( 
				\frac{\langle h \rangle }{ f } 
			\right)
				\equiv \frac{g^2 v^2}{4} 
\eeq
for the mass of the $W$ boson yields
\beq
	\frac{ v }{f} =
	\sin \left(	\frac{ \langle h \rangle }{ f} \right)
	=
	\sin \left(	\frac{ \sqrt{2} \langle \Phi \rangle }{ f }
				\right)
	\equiv
	s_{\alpha}	\;.
\label{eq:s_a}
\eeq

Higgs compositeness and the requirement for canonical 
normalization of the kinetic term lead to a rescaling of the 
physical Higgs field by a factor
$	c_{\alpha} = \sqrt{ 1- v^2/f^2}	$. 
This rescaling yields a reduction by a factor $c_{\alpha}$ of the 
couplings of the Higgs with the electroweak gauge bosons. 
The coupling of the Higgs boson to the fermions is also suppressed, 
but the suppression depends on how the fermions are embedded into 
representations of $SO(5)$. We will discuss this aspect in 
Section~\ref{sec:fermionic_sector}.

The reduction of the couplings between the Higgs and the gauge bosons 
leads to some sensitivity of the Peskin-Takeuchi $S$ and $T$ parameters~\cite{Altarelli:1990zd, Peskin:1991sw} 
on the cut-off $\Lambda$. In the SM the Higgs boson regulates the 
logarithmic divergencies of the gauge bosons self-energies.
The reduction of the Higgs boson couplings to the gauge boson spoils this effect.  
The result is a positive contribution 
to the $S$ parameter and a negative contribution to the $T$ parameter~\cite{Barbieri:2007bh}. 
The strong dynamics can further affect electroweak precision observables 
through some higher-order operator. Custodial symmetry is included to 
protect $T$, while the $S$ parameter receives a further positive contribution. 
We refer to~\cite{Barbieri:2007bh} for a more complete discussion of these effects.

The shifts to the $S$ and $T$ parameters from Higgs compositeness and 
from UV physics make the model incompatible with 
EWPT~\cite{Barbieri:2007bh, Gillioz:2008hs, Anastasiou:2009rv}. 
However, other composite states might lie below the cut-off of the effective 
theory and contribute to these observables. In particular, quark masses 
arise through mixing of the SM elementary fermions with fermionic 
bound states of the strong sector. We analyse this scenario and the effects on 
the electroweak parameters in the next Section.


\subsection{The fermionic sector}
\label{sec:fermionic_sector}

We extend the $SO(5)$ symmetry of the strong sector to the fermion sector 
and assume that composite top-partners lie below the cut-off of the effective 
theory description. Many viable choices for the embedding of these states into 
representations of $SO(5)$ have been considered, both in the context of 
five-dimensional and effective four-dimensional models~\cite{Agashe:2004rs, 
Contino:2006qr, Barbieri:2007bh, Falkowski:2007hz}. Following~\cite{Anastasiou:2009rv}, 
we include fermionic multiplets $\Psi$ that transform under the fundamental 
representation of $SO(5)$. Their $SO(5)$-invariant mass Lagrangian 
is 
\beq
\label{L:BSM}
-\lag_{SO(5)}= m^i_{\Psi}\bar{\Psi}^i \Psi^i
		+\mu_{ij} f (\bar{\Psi}^i \Sigma) (\Sigma^\dagger \Psi^j) \;, 
\eeq
where the indices $i,j$ allow for the possibility of more than one set
of fermionic composites and the brackets in the second term indicate
the contraction of the $SO(5)$ indices. In Eq.~(\ref{L:BSM}), 
$\mu_{ij}$ is a hermitian matrix 
and $\Sigma$ is the Higgs field defined in~(\ref{eq:sigma_expanded}). 

For models with one multiplet, only a small region of parameter space 
is allowed by electroweak precision measurements~\cite{Gillioz:2008hs, 
Anastasiou:2009rv}. 
The introduction of more sets of fermionic composites is inspired by 
the five-dimensional model presented in~\cite{Panico:2008bx}, which 
does not seem to suffer from such severe constraints. 
In~\cite{Anastasiou:2009rv}, it was shown that a four-dimensional 
composite Higgs model with two multiplets of fermionic resonances 
is compatible with EWPT in large regions of parameter space. 

A vector $(5)$ of $SO(5)$ decomposes under $SO(4) \sim SU(2)_L\times SU(2)_R$ 
into a bidoublet $(Q,X)$ plus a singlet $T$, 
\beq
	\Psi = (Q,X,T) 
	\Rightarrow 
	(5)= (4)\oplus (1) \simeq (2,2)\oplus (1,1) \;.
\eeq
We assign to $\Psi$ an $U(1)_X$ charge of 2/3. 
In this way the $SU(2)_L$ doublets $Q$ and $X$ 
get hypercharge $1/6$ and $7/6$, respectively, 
and the singlet $T$ acquires hypercharge 2/3. 
Therefore the SM quarks $q_L$ and $t_R$ have the same quantum numbers as 
$Q$ and $T$. We can write for them an interaction Lagrangian
\beq
\label{L:int}
	-\lag_{int}=m_L^{i} \bar{q}_L Q^i_R + m_R^{i}
									\bar{T}^i_L t_R + h.c. \;. 
\eeq
The doublet $X$ introduces a new quark of electromagnetic charge 
$2/3$, which mixes with the top after electroweak symmetry breaking, 
and an exotic quark of charge $5/3$ that does not couple to the 
Higgs boson. 

Using Eqs.~(\ref{L:BSM}) and~(\ref{L:int}) and expanding $\phi^0$ 
in~(\ref{eq:sigma_expanded}) around its vacuum expectation value, 
$\phi^0 = \frac{ v + h }{ \sqrt{ 2 }} $, 
we obtain the mass terms and the Yukawa couplings of the quarks. 
The mass matrix for the quarks of charge $2/3$ reads
\beq
\label{eq:mtop}
	- \mathcal{L}_m^{t}
	=
	\begin{pmatrix}
		\overline{t}_L \\ \overline{Q^u}_L \\ \overline{X^d}_L \\ \overline{T}_L 
	\end{pmatrix}^T
	\begin{pmatrix}
		0		&	m_L^\mathrm{T}										&	0																&	0																\\
		0		&	m_\Psi+s_\alpha^2 \frac{ f \mu }{2}		&	s_\alpha^2 \frac{ f \mu }{2}						&	s_\alpha c_\alpha \frac{f \mu}{\sqrt{2}}  \\
		0		&	s_\alpha^2 \frac{ f \mu }{2}						& m_\Psi+s_\alpha^2 \frac{ f \mu }{2}		&	s_\alpha c_\alpha \frac{f \mu}{\sqrt{2}}  \\ 
		m_R	&	s_\alpha c_\alpha \frac{f \mu}{\sqrt{2}}  & s_\alpha c_\alpha \frac{f \mu}{\sqrt{2}}  & m_\Psi + c_\alpha^2f \mu 
	\end{pmatrix}
	\begin{pmatrix}
		t_R \\ Q^u_R \\ X^d_R \\ T_R
	\end{pmatrix}
	+
	h.c. \;\;.
\eeq
The indices $u$ and $d$ denote the upper and lower components of a doublet, 
respectively.  In the case of more fermionic resonances, the mass matrix 
is to be understood in block form.

Since the mass of the bottom quark is small, we do not expect large effects from 
bottom compositeness. Instead of generating a bottom mass introducing 
additional $SO(5)$ multiplets, we adopt a minimal description 
and introduce an explicit $SO(5)$-breaking term
\beq
\label{eq:m_b}
	\mathcal{L}^b
	=
	\lambda_b \bar{q}_L \Phi b_R
\eeq
to give a mass to the bottom quark. 
Therefore the mass matrix for the quarks of charge -1/3 is
\beq
\label{eq:mbott}
	- \mathcal{L}_m^{b}
	=
	\begin{pmatrix}
		\overline{b}_L \\ \overline{Q^d}_L
	\end{pmatrix}^T
	\begin{pmatrix}
		- s_\alpha \frac{ \lambda_b f}{\sqrt{2}}	&	m_L^\mathrm{T} \\
		\quad 0	&	m_\Psi \\
	\end{pmatrix}
	\begin{pmatrix}
		b_R \\ Q^d_R
	\end{pmatrix}
	+
	h.c. \;\;.
\eeq
The Yukawa couplings of the top-like quarks are 
\beq
	-\mathcal{L}^t_h
	=
	h
	\begin{pmatrix}
		\overline{t}_L \\ \overline{Q^u}_L \\ \overline{X^u}_L \\ \overline{T}_L 
	\end{pmatrix}^T
	\left[
	\begin{pmatrix}
		0	&	0	&	0	&	0 \\
		0	&	s_{\alpha} c_{\alpha}	&	s_{\alpha} c_{\alpha}	&	\frac{1-2 s_{\alpha}^2}{ \sqrt{2} } \\
		0	& s_{\alpha} c_{\alpha}		&	s_{\alpha} c_{\alpha}	&	\frac{1-2 s_{\alpha}^2}{ \sqrt{2} } \\
		0  &	\frac{1-2 s_{\alpha}^2}{ \sqrt{2} }	&	\frac{1-2 s_{\alpha}^2}{ \sqrt{2} }	&	-2 s_{\alpha} c_{\alpha}
	\end{pmatrix}
	\otimes \mu
	\right]
	\begin{pmatrix}
		t_R \\ Q^u_R \\ X^u_R \\ T_R
	\end{pmatrix} 
	+ h.c.
	\;\;,
\eeq
where $\otimes$ denotes the matrix tensor product. 

The most important bounds on the model come from the $S$ and $T$ parameters and the anomalous 
$Zb_L \bar{b}_L$ coupling. We use the fit to these three quantities that was employed in~\cite{Anastasiou:2009rv}. 
The scan over the parameter space is done as in Ref.~\cite{Dissertori:2010ug}. 
We choose $f=500$~GeV, $m_H=120$~GeV, and we set the 
top and bottom masses to~\cite{Nakamura:2010zzi, Kuhn:2007tn} 
\beq
	 m_t = 172 {\rm \; GeV \qquad and \qquad} m_b = 4.16 {\rm \; GeV \;.}
\eeq

Direct experimental searches impose lower limits on the masses of the new quarks. 
These analyses however assume that the new quarks decay entirely through one 
specific channel 
($b' \to t W^{-}$~\cite{Aaltonen:2011vr, Chatrchyan:2011em}, 
$b' \to b Z$~\cite{Aaltonen:2007je}, 
$t' \to b W^{+}$~\cite{tpWb}, $t' \to q W^{+}$~\cite{tpWq}, 
$X \to t W^{+}$~\cite{Aaltonen:2009nr}). 
Studies carried out in the context of a four-generation Standard Model 
show that the bounds can be significantly lowered when multiple decay 
channels are open~\cite{Flacco:2010rg}. Following~\cite{Dissertori:2010ug}, 
we impose the limits
\beq
	m_{5/3} > 365 {\rm \; GeV \qquad, \qquad } m_{2/3},\, m_{-1/3} > 260 {\rm \; GeV }
\eeq
on the masses of the new quarks of charge $5/3$, $2/3$ and $-1/3$.



\section{Higgs production in composite Higgs models}
\label{sec:numerics}


\subsection{General LO results}
\label{subs:sigma_LO}
We first compute the contribution from the charge 2/3 quarks to the LO 
Higgs production in the heavy-mass approximation ($m_q > 2 m_H$). 
This is an interesting analysis, as we can prove that in this approximation 
the cross section is suppressed with respect to the SM result by a factor that only 
depends on the scales of the electroweak symmetry breaking $v$ 
and of the global symmetry breaking $f$. Such result is already 
known for the case of one multiplet~\cite{Falkowski:2007hz}. 
We show that it holds for any number of multiplets. 

Denoting by $\sigma_{app}^{CH(SM)}$ the LO production 
cross section in the heavy quark approximation 
in the composite Higgs (Standard) Model, the suppression factor 
reads 
\beq
\label{eq:LO_suppression}
	R_g = \frac{ \sigma_{app}^{CH} }{ \sigma_{app}^{SM} }
			= \left[ \frac{	\cos \left(2 \langle h \rangle /f \right)	}
								{	\cos \left( \langle h \rangle /f \right)	} \right]^2
			= \frac{	\left(1 - 2 s_{\alpha}^2 \right)^2 }{ 1 - s_{\alpha}^2}  \;.
\eeq
For our choice of parameters, $R_g = 35\%$. 

Our proof follows the one of Ref.~\cite{Falkowski:2007hz}.
In the heavy-mass approximation, the SM Higgs production amplitude 
$ {\cal M}_{gg \to H}^{SM}$ 
can be written as (Eqs.~(\ref{eq:M0ggh}), (\ref{eq:WCfinal}))
\beq
	 {\cal M}_{gg \to H}^{SM}
	 =
	 f(\ep, p) \frac{Y_{top}}{ m_{top} }
	=
	 f(\ep, p) \frac{ 1 }{ v } \;.
\eeq
In this expression, $f(\ep, p)$ contains the dependence on the polarization and momentum of 
the external gluons and $m_{top}, Y_{top}$ are the mass and Yukawa coupling 
of the top quark, respectively. 
In the presence of more heavy quarks of mass $m_i$ and Yukawa coupling $Y_i$, 
this result generalizes to 
\beq
	 {\cal M}_{gg \to H}
	 =
	 f(\ep, p) \sum_i \frac{Y_{i}}{ m_{i} } \;,
\eeq
so that
\beq
\label{eq:rg1}
	R_g^{1/2} 
		= v \sum_i \frac{Y_{i}}{ m_{i} } 
		= v \, {\rm Tr} \left( M^{-1} Y \right) \;.
\eeq
Here  $M$ and $Y$ denote respectively the matrices of masses and Yukawa couplings. 
Using  
\beq
	Y = \frac{ \partial M }{ \partial \langle h \rangle } \;,
\eeq
we can rewrite (\ref{eq:rg1}) as 
\beq
\label{eq:RgdetM}
	R_g^{1/2} 
		= v \, {\rm Tr} \left(		M^{-1}  \frac{ \partial M }{ \partial \langle h \rangle } 	 \right) 
		= v \frac{ \partial }{ \partial \langle h \rangle } \, {\rm Tr} \log M
		= v \frac{ \partial }{ \partial \langle h \rangle } \log \det M 
		\;.
\eeq
The dependence of the determinant of the mass matrix 
on 
$\langle h \rangle$ (i.e. on $s_{\alpha}, c_{\alpha}$) is of the form 
\beq
\label{eq:detmtext}
	\det M = s_{\alpha} c_{\alpha} \xi ( m_L, m_R, m_{\Psi}, \mu , f )  
	\;,
\eeq
where $\xi ( m_L, m_R, m_{\Psi}, \mu , f ) $ is a function of the 
parameters indicated. We derive the expression for $\det M$ and give the explicit 
form of the function $\xi$ in Appendix~\ref{app:detM}. 
Inserting~(\ref{eq:detmtext}) into~(\ref{eq:RgdetM}) and using the definition of 
$s_\alpha$~(Eq.~(\ref{eq:s_a})), we obtain
\beq
	R_g^{1/2} 
		= v \frac{ \partial }{ \partial \langle h \rangle } \log 
			\left[
				\sin \left( \frac{ \langle h \rangle }{f } \right)
				\cos \left( \frac{ \langle h \rangle }{f } \right)
				\xi
			\right]
			=
			\frac{\cos \left( 2 \langle h \rangle /f  \right)}
			{\cos \left( \langle h \rangle /f  \right)} \;.
\eeq
As anticipated, this result does not depend on any of the 
parameter-space details, 
including the number of fermionic multipets. 

Finite-mass corrections, bottom-quark and electroweak effects modify this result 
already at LO. 
In particular, bottom-quark and electroweak contributions are more significant than in the SM. 
In the SM, the inclusion of the bottom quark lowers the LO Higgs production cross section by about $7\%$, 
while electroweak effects give a $~5\%$ increase.  
As we have seen, in the composite Higgs model the contribution from heavy quarks is strongly suppressed. 
On the other hand, in our description we couple the bottom quark directly to the Higgs boson. As a 
consequence, its Yukawa coupling is 
reduced only by about $13\%$ with respect to the SM value\footnote{
	This suppression is related 
	to the mechanism that we adopt to give a mass to the bottom quark. 
	 In particular, it can be modified in a scenario where  an additional $SO(5)$ 
	 multiplet is introduced and the bottom quark 
	 acquires a mass in a similar way as the top quark.}. 
One therefore expects a larger reduction of the cross section from bottom quark loops, of the order of $10\%$. 
Similarly, the couplings of the gauge bosons to the Higgs are reduced by 
a factor $c_{\alpha} \sim 87\%$ with 
respect to their SM value. At LO in $\alpha_s$, the contribution from electroweak corrections should 
therefore yield an increase of the cross section by about $7\%$, against the $+5\%$ of the SM. 
These estimates are confirmed by the exact numerical values for the LO production cross section  that we 
report in Table~\ref{tab:LO_comparison}. 
\begin{center}
\begin{table*}[tb]
\begin{center}
\begin{tabular}{|c|c|c|c|c|c|c|}
\hline
	&	$\sigma^{LO}_t [pb]$
	&	$\sigma^{LO}_{tb} [pb]$	&	$\frac{\sigma^{LO^{\phantom{I}}}_{tb} \!\! -\sigma^{LO}_{t}}{\sigma^{LO}_{t_{\phantom{Q}}}}$
	&	$\sigma^{LO}_{te} [pb]$	&	$\frac{\sigma^{LO}_{te}-\sigma^{LO}_{t}}{\sigma^{LO}_{t}}$
	&	$\sigma^{LO} [pb]$
\\ 
\hline
	{\rm SM}	&	8.8		&
	8.1		&	--7\%	&
	9.2		&	+5\%	&
	8.6
\\
\hline   
	{\rm CH}	&	(2.9 $\div$ 3.2)		&
	(2.6	$\div$ 2.8)		&	--10\%	&
	(3.2	$\div$ 3.4)		&	 +7\%	&
	(2.9	$\div$ 3.1) 
\\
\hline  
\end{tabular}
\end{center} 
\caption{
  Leading-order gluon-fusion cross section in the 
  SM and in the composite Higgs model of Section~\ref{sec:composite_higgs_models} for the 7 TeV LHC. 
  The notation $(x_{min} \div x_{max})$ indicates the range of values that the quantity $x$ can assume. 
  We report the cross section $\sigma^{LO}_t$ due to charge 2/3 quarks only (including finite-mass effects), 
  and analyse how it changes with the inclusion of bottom-quark and electroweak corrections 
  ($\sigma^{LO}_{tb}$ and $\sigma^{LO}_{te}$). In the last column we give the total LO cross section $\sigma^{LO}$.}
\label{tab:LO_comparison}
\end{table*} 
\end{center}
%


\subsection{Precise prediction through NNLO}

We now compute the Higgs production cross section in the composite Higgs model through NNLO. 
We include the contribution from the heavy quarks retaining the full mass dependence through 
two loops. The NNLO corrections are computed in the effective theory approximation according to the 
Wilson coefficient~(\ref{eq:WCfinal}). Since all the heavy quarks are integrated out from the low-energy 
effective theory, the only difference with the SM calculation is in the expression of the Wilson coefficient. 
The remaining part of the 
calculation is the same as in the SM~\cite{Harlander:2002wh,Anastasiou:2002yz,Ravindran:2003um}. 
Following the approach of~\cite{Anastasiou:2008tj}, the NNLO corrections are normalized to the exact 
LO cross section according to
\begin{equation}
	\sigma^{NNLO;heavy} 
	\simeq 
	\sigma_{exact}^{LO;heavy} \cdot
	\left(  
		\frac{ \sigma^{NNLO;heavy} }{ \sigma^{LO;heavy}} 
	\right)_{effective} 
	\;.
\end{equation}
Since bottom-quark effects are more important than in the SM, we compute them exactly 
through NLO~\cite{Anastasiou:2009kn,Spira:1995rr}. 
We also include the full two-loop SM electroweak corrections of Ref.~\cite{Actis:2008ug} and the 
three-loop mixed QCD and electroweak corrections derived within an effective-theory approach 
in Ref.~\cite{Anastasiou:2008tj}. 
Both the electroweak corrections are rescaled by the factor $c_\alpha$ that reduces the coupling 
of the Higgs to the gauge bosons. 
All the effects described here are included in the code~\texttt{iHixs}~\cite{iHixs11}, which we 
use for the calculation of our results. 
\begin{center}
\begin{table*}[b]
\begin{center}
\begin{tabular}{|c|c|c||c|c|c||c|}
\hline
		$\sigma^{SM^{\phantom I}} [pb]$
	&	$\delta^{(+)}_{\rm scale}$ \%	
	&	$\delta^{(-)}_{\rm scale}$ \%	
	&	$\sigma^{CH} [pb]$	
	&	$\delta^{(+)}_{\rm scale}$ \%	
	&	$\delta^{(-)}_{\rm scale}$ \%	
	&	$R'_g$
\\ 
\hline
		17.6						&			+9	\%		& 		--10	\%		&
		(5.9 $\div$ 6.4)		&	+(6 $\div$ 12)\%	&	--(7 $\div$ 11)\%&
		(34 $\div$ 37) \%
\\
\hline  
\end{tabular}
\end{center} 
\caption{
  Gluon-fusion cross section through NNLO in the 
  SM and in the composite Higgs model, 
  with 
  the corresponding scale variation errors. The factor $R'_g$ is defined as in
  Eq.~(\ref{eq:LO_suppression}), but for the full result through three loops.}
\label{tab:NNLO_results}
\end{table*} 
\end{center}
In Table~\ref{tab:NNLO_results} we present the full NNLO cross section in the SM and in the composite 
Higgs model. The results are similar for the case of one and of two multiplets of composite fermions. 
We  use  the MSTW2008 NNLO parton distribution 
functions~\cite{Martin:2009iq} and set the renormalization and factorization scales to 
$\mu = \mu_f=\mu_r = m_{H}/2$. 
The $35\%$ suppression factor $R_g$ computed in Section~\ref{subs:sigma_LO} is confirmed through NNLO. 
We estimate the uncertainty due to higher order corrections by varying the scale $\mu$ in the interval 
$\left(m_H/4, m_{H}\right)$. This scale variation uncertainty is similar to the SM one. 

Finally, in Table~\ref{tab:kfactors} we compare the K-factors in the SM and in the composite Higgs 
model. As in the SM, the K-factors are large and the NNLO result is about twice as big as the LO 
cross section.
\begin{center}
\begin{table*}[tb]
\begin{center}
\begin{tabular}{|c|c|c|}
\hline
	&				SM			&			CH
\\ 
\hline
	$	\frac{\sigma^{NLO^{\phantom I}}}{\sigma^{LO}_{\phantom I}}$		&
	+ 75\%			&
	+ (77 $\div$	 78) \%
\\ 
\hline
	$	\frac{\sigma^{NNLO^{\phantom I}}}{\sigma^{LO}_{\phantom I}}$		&
	+ 106\%			&
	+ (108 $\div$	 110) \%
\\
\hline  
\end{tabular}
\end{center} 
\caption{
  NLO and NNLO K-factors in the SM and in the composite Higgs model.}
\label{tab:kfactors}
\end{table*} 
\end{center}
%



\section{Conclusions}
\label{sec:conclusions}

We presented the construction of an effective theory for  extensions  of the Standard Model with 
an arbitrary number of heavy quarks coupling to the Higgs. We assumed a general form for the 
Yukawa couplings of these quarks to the Higgs boson. This situation arises for example in 
the context of composite Higgs model, where the mass of the quarks can be explained  through 
the mixing of the fundamental SM particles with heavy composite fermions.  
We computed the  Wilson coefficient of the effective Higgs-gluon vertex 
through ${\cal O}(\alpha_s^3)$. 
We used our result to compute the Higgs production cross section through NNLO in a 
composite Higgs model with an $SO(5) \to SO(4)$ global symmetry breaking pattern and  
one or two multiplets of composite fermions transforming under the fundamental representation 
of $SO(5)$. We showed that, in the heavy quark-mass approximation, the LO production cross 
section is suppressed with respect to the SM value by a factor that depends neither on the 
details of the parameter space nor on the number of multiplets. 
As in the SM, the NNLO result is enhanced with respect to the LO cross section by approximately a 
factor of 2. The scale variation errors also behave in a similar way as in the SM. 
We included in our result also the full dependence on the bottom quark mass through two loops 
and the two-loop electroweak and three-loop mixed QCD and electroweak corrections. 
Both these effects are enhanced with respect to the SM. As in the SM, they give contributions 
of opposite sign, which cancel. 

In this work we applied our result for the Wilson coefficient to a 
specific beyond-the Standard Model scenario, but the calculation can 
be extended to any model with additional new quarks in the 
fundamental representation of the colour group.



\section*{Acknowledgements}
We thank Babis Anastasiou and Giuliano Panico for many useful discussions  
and for their comments on the script, 
and Achilleas Lazopoulos for providing the preliminary version of~\texttt{iHixs}. 
We greatly appreciated the hospitality of the ETH theory group 
during parts of this work.
This research is supported by the DOE under Grant DE-AC02-98CH10886.



\appendix
\section{Analytical form of $\det M$}
\label{app:detM}

We derive here the analytical result for the determinant of the mass matrix of charge 2/3 quarks, $\det M$, 
for an arbitrary number of multiplets. 
Let us recall (Eq.~(\ref{eq:mtop})) that the mass matrix for the charge 2/3 quarks reads 
\beq
M
=
	\begin{pmatrix}
		0		&	m_L^\mathrm{T}										&	0																&	0																\\
		0		&	m_\Psi+s_\alpha^2 \frac{ f \mu }{2}		&	s_\alpha^2 \frac{ f \mu }{2}						&	s_\alpha c_\alpha \frac{f \mu}{\sqrt{2}}  \\
		0		&	s_\alpha^2 \frac{ f \mu }{2}						& m_\Psi+s_\alpha^2 \frac{ f \mu }{2}		&	s_\alpha c_\alpha \frac{f \mu}{\sqrt{2}}  \\ 
		m_R	&	s_\alpha c_\alpha \frac{f \mu}{\sqrt{2}}  & s_\alpha c_\alpha \frac{f \mu}{\sqrt{2}}  & m_\Psi + c_\alpha^2f \mu 
	\end{pmatrix} \;.
\eeq
Successively taking linear combinations of lines/columns of $M$, we can recast it into the form
\beq
M'
=
	\begin{pmatrix}
		\phantom{aa} m_L^\mathrm{T} \phantom{aa}	&	\phantom{aa}  0 \phantom{aa} 	&	\phantom{aaa}  0 \phantom{aaa} 	&	\phantom{aaaa}  0 \phantom{aa}  \\
		0																		&	m_R							&	- \frac{ s_{\alpha}^2 + 2 c_{\alpha}^2 }{ s_{\alpha} c_{\alpha}} \frac{m_\Psi}{ \sqrt{2}}	& m_\Psi \\
		s_\alpha^2 \frac{ f \mu }{2} 							&	0								& \;\;	 m_\Psi															&	s_\alpha c_\alpha \frac{f \mu }{\sqrt{2}} \\
		m_{\Psi}															&	0								&	-m_{\Psi}																&	0
	\end{pmatrix} 
	\equiv
	\begin{pmatrix}
		 { \bf A} & { \bf B} \\ { \bf C} & { \bf D}
	\end{pmatrix}
	\;,
\eeq
where
\bea
	{\bf A} =
	\begin{pmatrix}
		\phantom{a}  m_L^\mathrm{T}	\phantom{a}  & \phantom{a}  0 \phantom{aa} 	 \\
		0	&	m_R	
	\end{pmatrix} 
	&\quad, \quad &
	{\bf B} =
	\begin{pmatrix}
		\phantom{aa}  0 \phantom{aa} 	& \phantom{aa}  0	\phantom{a}  \\
		- \frac{ s_{\alpha}^2 + 2 c_{\alpha}^2 }{ s_{\alpha} c_{\alpha}} \frac{m_\Psi}{ \sqrt{2}}	& m_\Psi
	\end{pmatrix} 
	\qquad, \qquad \nn 
	{\bf C} =
	\begin{pmatrix}
	\phantom{a}  s_\alpha^2 \frac{ f \mu }{2} 	\phantom{a}  & \phantom{a} 	0 \phantom{a} 	 \\
		m_{\Psi}	&	0	
	\end{pmatrix} 
	&\quad {\rm and} \quad &
	{\bf D} =
	\begin{pmatrix}
		\quad \,  m_\Psi	\phantom{a}  &	\phantom{a}  s_\alpha c_\alpha \frac{f \mu }{\sqrt{2}} \phantom{a}  \\
			-m_{\Psi}	&	0
	\end{pmatrix} 
	\qquad .
\eea
Because of the properties of determinant, 
\bea
\label{eq:detmapp}
	\det M &=& \det M' =
	\det {\bf D} \det \left( {\bf A} - {\bf B} {\bf D}^{-1} {\bf C} \right) \nn
	&=&
	s_{\alpha} c_{\alpha} 
	\left[
		\frac{ 1 }{ \sqrt{2} } m_L^{T} W^{-1} m_R \det(m_\Psi) \det (f \mu) \det(W)
	\right] \;,
\eea
where
\beq
	W = m_{\Psi} + \frac{ 1 }{ f } m_{\Psi} \mu^{-1} m_{\Psi} \;.
\eeq
The quantity in square brackets in Eq.~(\ref{eq:detmapp}) corresponds to the 
function $\xi$ introduced in Eq.~({\ref{eq:detmtext}). 




\end{document}